\documentclass[a4paper]{jpconf}
\usepackage{graphicx}
\usepackage{amsmath,amssymb,amsfonts,amsthm,latexsym} 
\usepackage[pdftex,colorlinks=true,urlcolor=blue,linkcolor=blue,citecolor=blue,a4paper]{hyperref}
\usepackage{color}
\usepackage{float}
\usepackage{caption}
\usepackage{subcaption}

\begin{document}

\title{Dust in the Radio Galaxy and Merger Remnant NGC 1316 (Fornax A)
}

\author{B D Asabere$^1$, C Horellou$^2$, H Winkler$^1$, T Jarrett$^3$ and L Leeuw$^4$ }

\address{$^1$ Department of Physics, University of Johannesburg, P.O. Box 524, 2006, Auckland Park, Johannesburg, South Africa}
\address{$^2$ Department of Earth and Space Sciences, Chalmers University of Technology, Onsala Space Observatory, SE-439 92 Onsala, Sweden}
\address{$^3$ Department of Astronomy, University of Cape Town,  Private Bag X3,  Rondebosch 7701, South Africa}
\address{$^4$ Department of Interdisciplinary Research, University of South Africa, Pretoria, South Africa}

\ead{ bd.asabere@gmail.com}

\begin{abstract}
We present dust maps of NGC~1316 (Fornax~A), a well-studied early-type galaxy located in the outskirts of the Fornax cluster. We used the Large APEX BOlometer CAmera (LABOCA), operating at 870~$\mu$m with an angular resolution of 19.$''$5 on the Atacama Pathfinder EXperiment (APEX) 12~m submillimeter telescope in Chile and the Wide-field Infrared Survey Explorer ({\it WISE}). {\it WISE} observes in four mid-infrared bands centered at 3.4, 4.6, 12 and 22~$\mu$m with angular resolutions ranging from 6 to 12$''$. The {\it WISE} and LABOCA maps reveal emission from dust in the central 2$'$ of NGC~1316. The disturbed optical morphology with many shells and loops, the complex distribution of molecular gas 
and our dust maps are evidences of past merger activity or gas accretion in the galaxy. Combining the LABOCA flux measurement with existing mid- and far-infrared measurements, we estimate the temperature of the cold ($\sim$20 K) and warm ($\sim$55 K) dust components in the galaxy. This study will be extended to other southern radio galaxies and merger remnants. Those galaxies are good targets for future observations at higher angular resolution and sensitivity with {\it ALMA} to probe the interaction of the radio jets with the dusty molecular gas near active galactic nuclei.

\end{abstract}

\section{Introduction}
Dust is ubiquitous in the interstellar medium (ISM) of galaxies but it 
constitutes only about 1\% of the total mass of the ISM. Notwithstanding, dust plays an important role in galaxies: it is a tracer of star formation and stellar evolution, and contributes to the evolution of galaxies \cite{Spitzer78, Blain2002}. 
Interstellar dust grains span a wide range of sizes ({$\sim$0.001 - 0.25~$\mu$m}) and temperatures ($\sim$20 - 200~K). 
Dust causes extinction of starlight: it dims and reddens the galaxy light at ultraviolet (UV) and optical wavelengths, but 
re-radiates about 90\% of the absorbed galaxy energy into the infrared and submillimeter wavebands. 
Interstellar dust can thus be observed in different galactic environments by mapping the re-radiated emission at mid-infrared, far-infrared and submillimeter wavelengths \cite{Draine2009, Kennicutt2003}. 

\medskip
Contrary to the belief that early-type galaxies are ``red and dead" with little interstellar medium and star formation \cite{Faber76, Thomas2005}, recent studies have revealed that more than 30\% of nearby early-type galaxies are rich in molecular clouds, gas and dust \cite{Leeuw2008, Kuntschner2010, Bureau2011} due to past merger events that contribute to interstellar and nuclear activities.

\medskip

NGC~1316 (Fornax~A, PKS 0320-37, Arp~154) is a peculiar dusty early-type radio galaxy located in the 
outskirts of the Fornax cluster; roughly 2~Mpc from the core of the cluster and at an adopted distance of 21.0~Mpc \cite{Blakeslee2009}. It is one of the two most luminous galaxies in the Fornax cluster and the third most powerful nearby radio source besides Centaurus~A and Virgo~A. It has a pair of radio lobes lying outside the optical galaxy \cite{Ekers83} and a two-sided radio jet in the central 30$''$ \cite{Geldzahler84}. The galaxy has a disturbed outer morphology with numerous loops, ripples, tidal tails and shells \cite{Schweizer88, Schweizer80, Matthews64}. Spectroscopy of the brightest globular clusters suggests that a major merger took place about 3~Gyr ago \cite{Goudfrooij2004}. Other studies \cite{Schweizer80, Mackie98} peg a merger event about 0.5~Gyr ago. The abundance of molecular gas and the different kinematics of the stars and the gas \cite{horellou2001, Bosma85} support the idea of  a recent merger with a companion gas-rich galaxy.

\medskip  
NGC~1316 displays prominent dust patches in the central 2$'$.4 x 2$'$.4, with an inner dust lane of about 2$'$ oriented along the apparent optical minor axis at a position angle of $\sim$~140$^\circ$ \cite{Grillmair99, Lanz2010}. Massive dust-heating stars are being formed in the galaxy \cite{Temi2005}. The amount of dust estimated from optical extinction measurement is $\sim$ {$2.13\times10^5 M_\odot$}, that from {\it IRAS} total flux densities is {$2.11\times10^6 M_\odot$} and the estimate from the integrated flux densities at {\it Spitzer/MIPS} is {$3.2\times10^6 M_\odot$} \cite{Dale2007, Draine2007}. CO emission has been detected in two regions ($\sim$ 30$''$ southeast and $\sim$45$''$ northwest) of the nucleus, with corresponding molecular hydrogen masses of {$0.9\times10^8 M_\odot$} and {$2.2\times10^8 M_\odot$} respectively \cite{horellou2001}.  X-ray observations in the energy range 0.3 - 8.0~keV have revealed a low-luminosity AGN (${L_{X}\sim 5.0\times10^{39} erg~s^{-1}}$) \cite{kim2003, feigelson1995, fabbiano1992}. 
AGN feedback may inhibit star formation \cite{hopkins2010}. NGC~1316 is therefore a self-contained laboratory for studying galaxy formation and evolution, gas settling toward equilibrium after a merger activity, and feedback from AGN on the star-forming ISM.

\section{Observations and Data Processing}
NGC~1316 was observed in June 2012 for 8~hours and 10~minutes with LABOCA \cite{siringo2009}, a bolometer camera operating on the APEX 12~m submillimeter telescope \cite{gusten2006} at 5105~m elevation in the Atacama desert in Chile, one of the driest places on Earth. LABOCA observes in the continuum at a central wavelength of 870~$\mu$m ({345~GHz}) in total-power scanning observing mode. The angular resolution is 19.$''$5 and the field of view is 11.$'$4. Its passband has a full width half maximum of about 60~GHz ({$\sim$150~$\mu$m}) to match the corresponding atmospheric window. The version 2.12-2 of {\tt Crush-2} \cite{kovacs2008} was used to reduce the data and produced the full and detailed map of the dust. {\tt Crush-2} a is comprehensive bolometric data reduction utility and imaging software package for ground-based telescopes. Iteratively, {\tt Crush-2} removed correlated noise from the raw data in the digitized time-streams, identified and flagged problematic data pixels and provided clean  and independent bolometer signals in the individual ($\sim$70) scans, which were then co-added to produce the final maps. 
The flux measurement, signal-to-noise ratio and the root-mean-square analyses and estimations were all done with {\tt Crush-2}. 

\medskip
{\it WISE} (Wide-field Infrared Survey Explorer: \cite{wright2010}) is a space infrared telescope that mapped the whole sky in four mid-infrared bands at central wavelengths of 3.4, 4.6, 12 and 22~$\mu$m labeled $W1$ to $W4$. The angular resolutions are 6.1, 6.4, 6.5 and 12$''$ respectively \cite{jarrett2012}. The {\it WISE} All-sky Data Products and Atlas Images are archived by the NASA IPAC (Infrared Processing and Analysis Center) in the IRSA (Infra-Red Science Archive). However, reduced drizzled and cleaned {\it WISE} images of NGC~1316 in the four bands were used for the study (see \cite{jarrett2012, jarrett2013}). The {\tt ellipse} task in the IRAF (Image Reduction and Analysis Facility) isophote package was used to extract the flux information in the respective {\it WISE} bands following the guidelines provided in the {\it WISE} Explanatory Supplement \cite{jarrett2013, cutri2012}. 

\medskip

\section{Results and Discussion}
As one moves to longer wavelengths ($\lambda > 2~\mu$m), stars contribute less and less to the observed emission, and the ISM dust component begins to dominate.
The $W1$ and $W2$  bands (3.4 and 4.6 $\mu$m) are nearly extinction-free bands that are mostly sensitive to the evolved stellar population: the Rayleigh-Jeans part of the black-body emission of cool stars ($T > 2000$~K). The emission in $W3$ is dominated by polycyclic aromatic hydrocarbons (PAHs) at 11.3~$\mu$m and warm continuum dust emission from small grains. 
The $W4$ band (22 $\mu$m) is more sensitive to the warm dust emission, which comes from photon-dominated regions (or ultraviolet radiation fields) in the ambient ISM. 
LABOCA, at 870~$\mu$m, is sensitive to cold dust that tends to dominate the bolometric luminosity in star-forming galaxies.

\subsection{Flux measurements}
The measured fluxes are given in Table~\ref{Measurements}. The values measured  
in the four {\it WISE} bands are comparable to those measured in the respective Spitzer bands (at 3.6, 4.5, 8 and 24~$\mu$m; see \cite{Lanz2010}) for our chosen elliptical apertures. The {\it WISE} flux measurements were obtained by performing elliptical aperture photometry with the {\tt ellipse} task in the IRAF isophote package to get the total counts in each band (in data number (DN) units) after background subtractions, and applied the respective band corrections \cite{wright2010, jarrett2013}. The associated  errors were estimated from the corresponding uncertainty maps \cite{cutri2012}. The measured flux in each {\it WISE} band came from an ellipse centered on the image (with position coordinates ({03h~22m~41.7s, $-$37$^\circ$~12$'$~30$''$})). 
Both the ellipticities and position angles were held constant during the elliptical isophotal fittings (see Table~\ref{table}). The LABOCA flux was estimated in a circular aperture of about $2'$.

\subsection{Dust maps}
\begin{figure}[!tb]     
        \centering
        \begin{subfigure}[b]{0.42\textwidth}
                \includegraphics[width=1.3\textwidth]{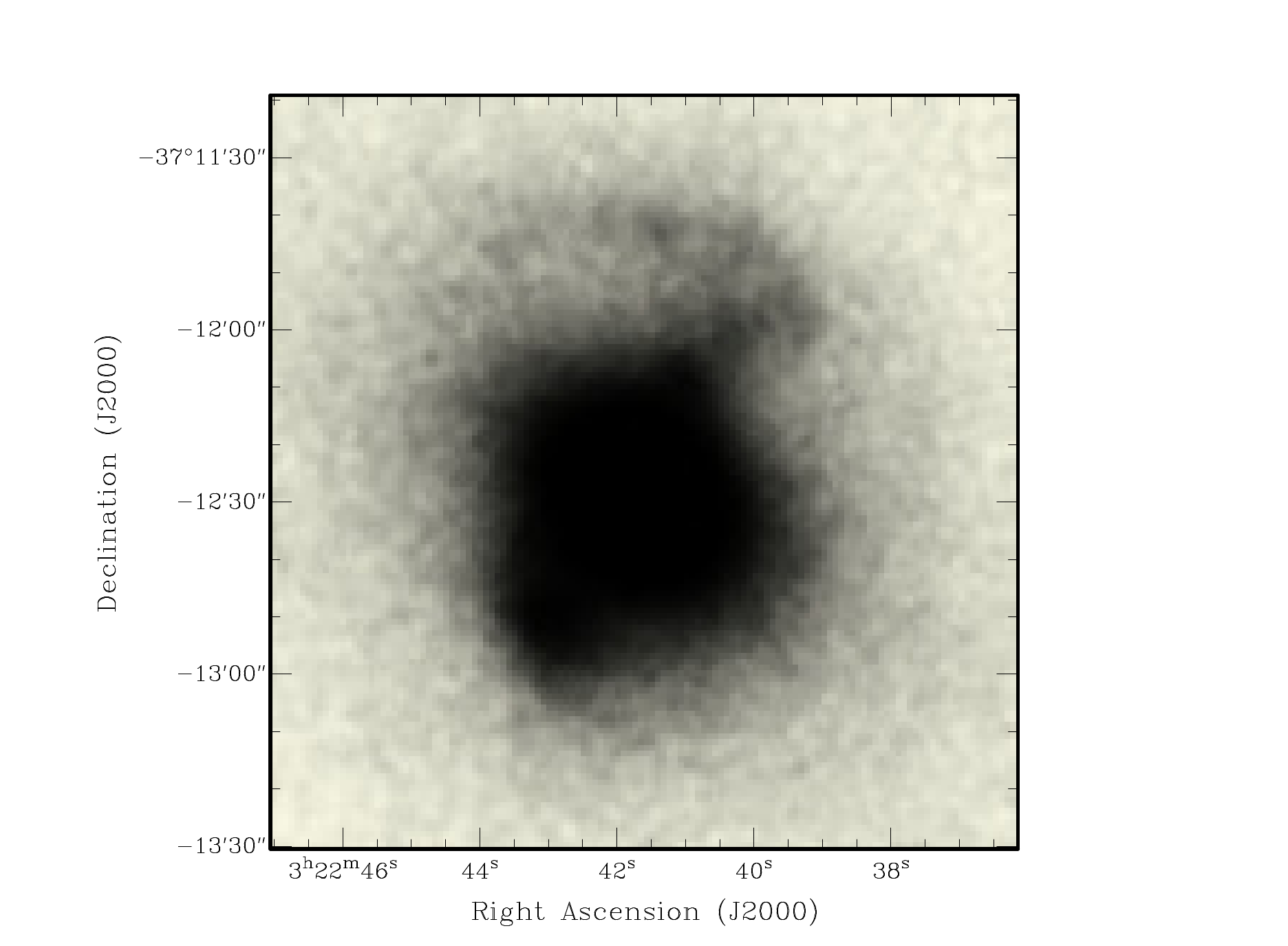}
                \caption{ W3 (12~$\mu$m) image}
                \label{image}
        \end{subfigure}
        \begin{subfigure}[b]{0.42\textwidth}
                \includegraphics[width=1.3\textwidth]{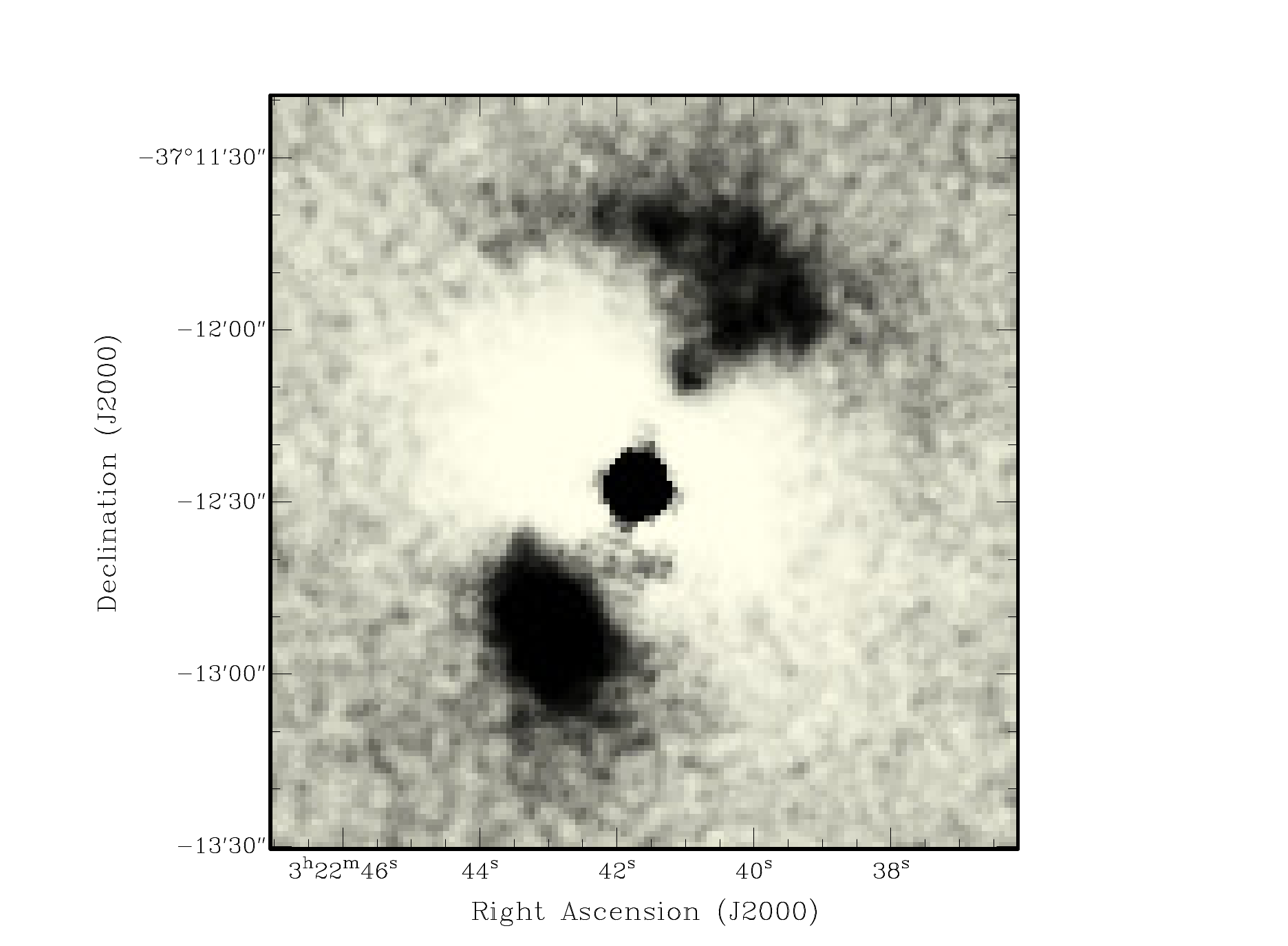}
                \caption{$W3$ dust map}
                \label{dustmaps}
        \end{subfigure}
        \caption{ {\it (a)} W3 (12~$\mu$m) map of NGC~1316 and
{\it (b)} the $W3$ dust map of the galaxy obtained after  
subtraction of the stellar component modeled as a S\'ersic model and 
sky component. The northern dusty ``arc" and the southern dust
concentration seen in the {\it HST} and in the {\it Spitzer} images \cite{Lanz2010} are 
apparent.}
\label{one}
\end{figure}

\begin{figure}[tb!]   
        \centering
        \begin{subfigure}[b]{0.42\textwidth}
                \includegraphics[width=1.24\textwidth]{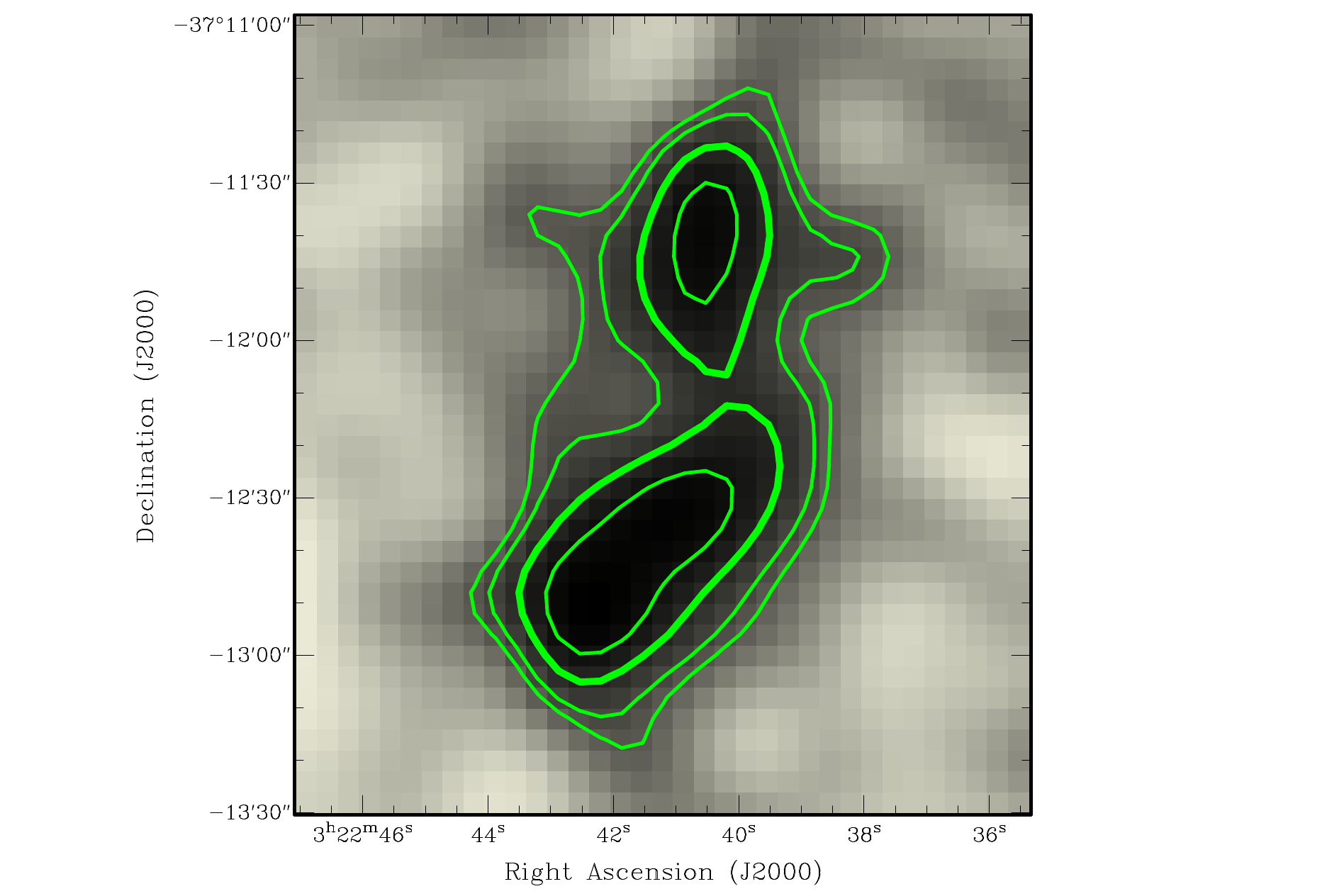}
                \caption{LABOCA dust map}
                \label{laboca}
        \end{subfigure}
        \begin{subfigure}[b]{0.42\textwidth}
                \includegraphics[width=1.24\textwidth]{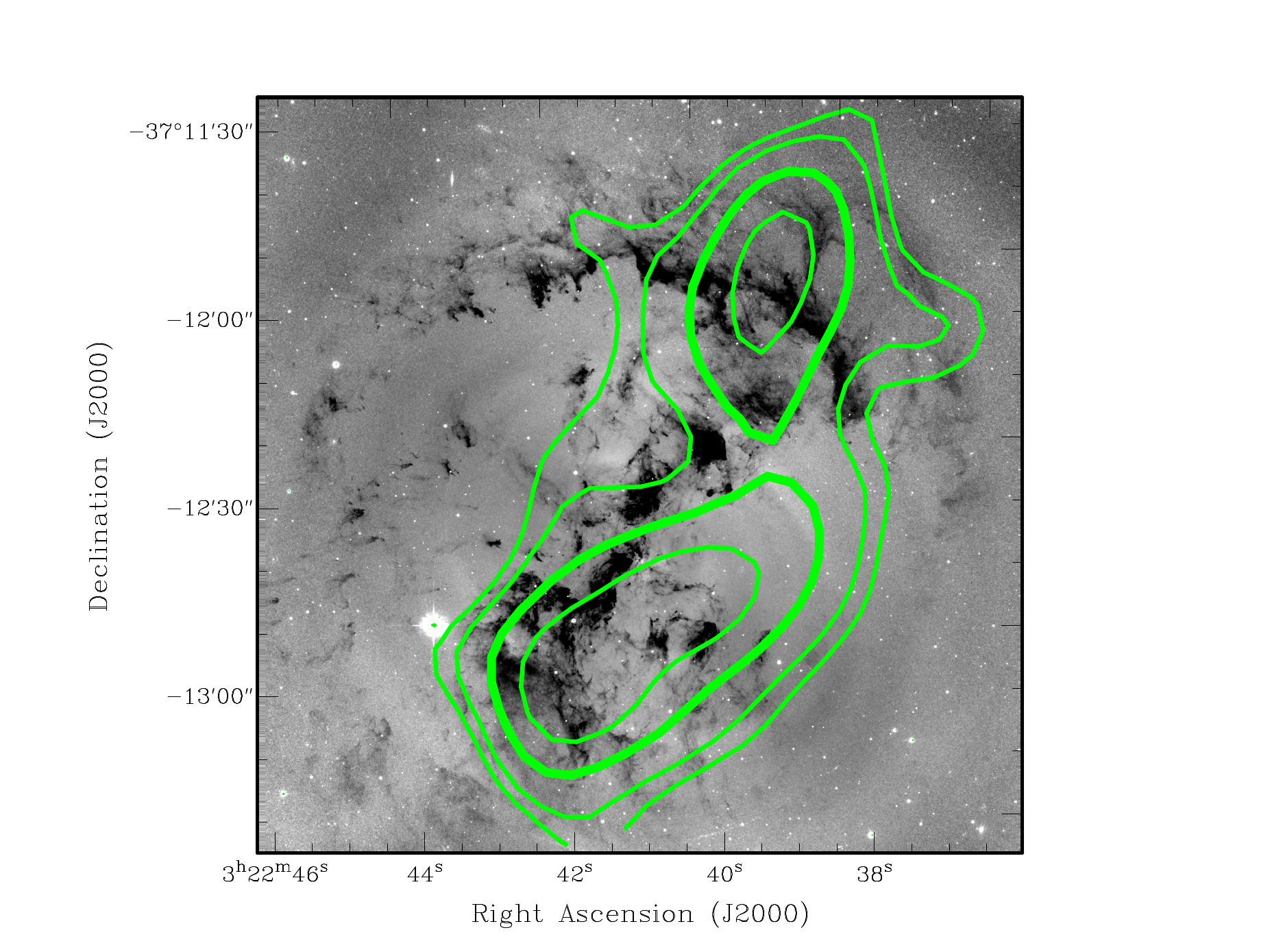}
                \caption{LABOCA dust map on {\it HST} image}
                \label{hst}
        \end{subfigure}
        \caption{{\it (a)} LABOCA 870~$\mu$m image of NGC~1316 in green contours overlaid on the greyscale image.
The contours range from 3 to 6 times the noise level, which is equal to 2.5~mJy/beam. The angular resolution of the image is 28$''$.  Figure {\it (b)} the {\it HST} image in grey-scale and LABOCA image in green contours. The two peaks detected in the submillimeter map correspond roughly to the dust lanes seen in extinction in the HST image.}
\label{dustcontours} 
\end{figure}

We used the software {\tt Galfit} \cite{peng2002} to identify the stellar component. We perform a two-dimensional fit of the superposition of a S\'ersic model, a point-source and sky background to the $W1$ image, following the same method as \cite{Lanz2010} in their analysis of {\it Spitzer} images. We then fixed all the parameters of the S\'ersic model from the $W1$ fit except the amplitude and center, and did a fit to the $W3$ image (see Figure~\ref{image}). The $W3$ dust map of the galaxy, shown in Figure~\ref{dustmaps}, was obtained by subtracting the S\'ersic model (identified as the stellar contribution) and the sky component. LABOCA dust contour map is shown in Figure~\ref{dustcontours}. The images in Figures~\ref{one} and \ref{dustcontours} were generated using the {\tt kvis} image display and manipulation program \footnote{{\url{http://www.atnf.csiro.au/computing/software/karma}}}.

\begin{table}[h!]
\caption{\label{Measurements} Flux measurements. The first column lists the four {\it WISE} bands and LABOCA with their corresponding central wavelengths in parenthesis. In the elliptical isophotal fits, fluxes were measured within ellipses with {\it semi-major axes (SMA)} lengths shown in column~2 at position angle of 34.7$^\circ$ for the {\it WISE} bands. The LABOCA flux was estimated in a circular aperture of about $2'$. The axis ratios listed in column~3 were derived by dividing the semi-minor axis by the semi-major axis lengths. The effective radii in columns~4 were also obtained from the isophotal fits with the {\tt ellipse} task in the isophote package of IRAF. The total counts in each elliptical aperture were converted into integrated fluxes, given in column~5, after extended aperture corrections \cite{jarrett2013, cutri2012}. }

\begin{center}   
\begin{tabular}{lllll}
\br
Band &SMA&Axis ratio&Effective radius&Flux density\\
 &(arcsec)&&(arcsec)&(Jy)\\
\mr
W1 (3.4 $\mu$m)&766.6&0.645&103.3&2.870$\pm$0.031\\
W2 (4.6 $\mu$m)&586.8&0.645&94.5&1.501$\pm$0.016\\
W3 (12 $\mu$m)&187.9&0.645&60.5&0.451$\pm$0.005\\
W4 (22 $\mu$m)&101.6&0.645&39.4 &0.342$\pm$0.004\\
LABOCA (870~$\mu$m) 	& -    &-& -	&0.113$\pm$0.015\\
\br
\end{tabular}
\end{center}
\label{table}
\end{table}

\subsection{Spectral energy distribution}
\begin{figure}    
\centering
\includegraphics[width=7.8cm,height=6.3cm]{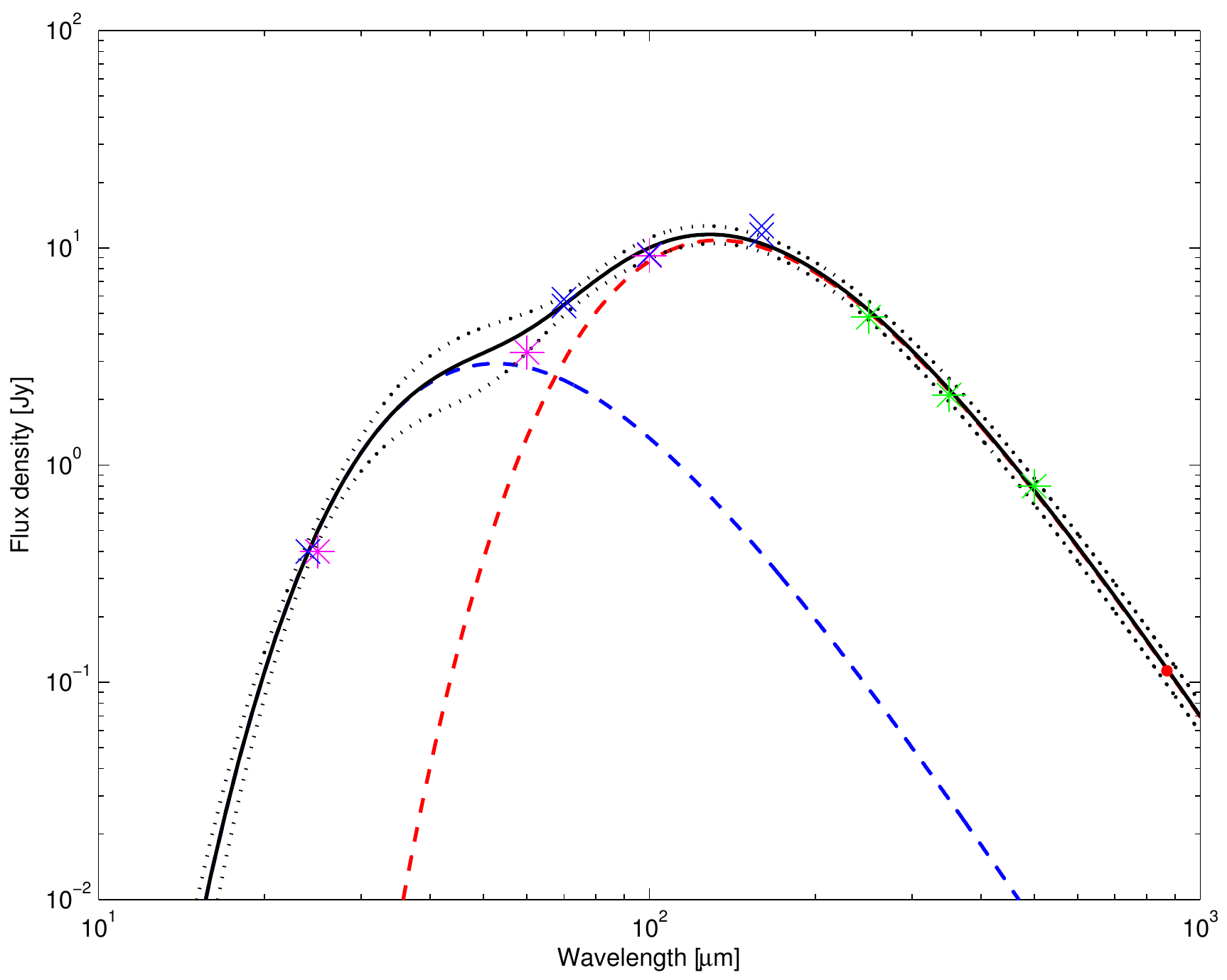}
\caption{\label{seds}Mid-infrared to submm spectral energy distribution of NGC~1316. 
The points are the measurements listed in Table~2 of \cite{galametz2012}  and our LABOCA point shown in red. The magenta stars are from {\it IRAS}, the blue crosses from {\it Spitzer} (refer to \cite{Lanz2010}) and the green stars from {\it Herschel} (see \cite{galametz2012}). The dashed lines show the best-fit models of the cold dust emission (in red) and the warm dust emission (in blue). The black solid line is the sum of the two, and the dotted lines show the 95\% confidence interval. The emissivity index was fixed to 2 for both components. }
\label{sedfit}
\end{figure}

We did a fit to published mid-infrared flux measurements and to our LABOCA submillimeter measurement, to model the SED of NGC~1316, as shown in Figure~\ref{sedfit}. We modeled the dust emission as the superposition of two modified black-body components, at different temperatures, 
\begin{equation}
S_{\nu} = A_{w} \lambda^{-\beta_w} B_{\nu}(T_{w}) + A_{c} \lambda^{-\beta_c} B_{\nu}(T_{c})
\label{b}
\end{equation}
 where ${\beta_w}$ and ${\beta_c}$ are the dust emissivity indices for the respective warm ($T_{w}$) and cold ($T_{c}$) temperature components. We fixed the emissivity indices to 2. 
$B_{\nu}(T)$ is the Planck function at the temperature of interest, $A_{w}$ and $A_{c}$ are amplitudes  \cite{galametz2012}. 
The inferred temperature values are $T_w = 55.0\pm4.2$~K and $T_c = 21.7\pm1.3$~K, in agreement with previous estimates \cite{galametz2012}.

\newpage
\section{Conclusion}
We have presented an ongoing analysis of the dust emission in the galaxy NGC~1316 based on 
{\it WISE} and LABOCA measurements in the mid-infrared and the submillimeter wavelengths. The $W3$ dust map revealed in this study, represents the PAHs (at 11.3~$\mu$m) dust component of the galaxy, which is shown for the first time in this galaxy. The central region of the $W3$ dust map appears over-subtracted. Separation of the stellar component from the dust is crucial and will be refined. Recently, \cite{galametz2014} presented LABOCA observations of NGC~1316. Our 870~$\mu$m flux measurement is lower than theirs but consistent within the error bars. A more detailed analysis of the $WISE$ and LABOCA images will be presented in a forthcoming paper. NGC~1316 is a target of choice for future observations of the dust and molecular gas with {\it ALMA} to study their possible interaction with the inner radio jet.

\newpage
\section{Acknowledgements}
We thank Zolt Levay and Paul Goudfrooij for making their {\it HST} image available to us.



\section*{References}
\begin{thebibliography}{35}
\bibitem{Spitzer78} 
Spitzer L 1978 {\it Physical Processes in the Interstellar Medium} (New York: Wiley)  pp 250--333 
\bibitem{Blain2002}
Blain A W,  Smail I, Ivison R J, Kneib J P and Frayer D T 2002 {\it Phys Reports} Issue 2 {\bf 369} 111--76
\bibitem{Draine2009}
Draine B T 2009 {\it Space Sci. Revs.} {\bf 143} 333--45
\bibitem{Kennicutt2003} 
Kennicutt R C {\it et al.} 2003 {\it Publ. Astron. Soc. Pac.} {\bf 115} 928--52
\bibitem{Faber76}
Faber S M and Gallagher J S 1976 {\it Astrophysical J.} {\bf 204} 356
\bibitem{Thomas2005}
Thomas D, Maraston C, Bender R and Mendes de Oliveira C 2005 {\it Astrophysical J.} {\bf 621}  673--94
\bibitem{Leeuw2008}
Leeuw L L, Davidson J, Dowell C D and Matthews H E 2008 {\it Astrophysical J.} {\bf 677} 249--61
\bibitem{Kuntschner2010}
Kuntschner H {\it et al.} 2010 {\it Mon. Not. R. Astron. Soc.} {\bf 408} 97--132
\bibitem{Bureau2011}
Bureau M {\it et al.} 2011 {\it IAU Symposium Proceedings} {\bf 277} 55--8
\bibitem{Blakeslee2009}
Blakeslee J P {\it et al.} 2009  
{\it Astrophysical J.} {\bf 694} 556--72
\bibitem{Ekers83}
Ekers R D, Goss W M, Wellington K J, Bosma A, Smith R M and Schweizer F 1983 {\it A \& A} {\bf 127} 361--65
\bibitem{Geldzahler84}
Geldzahler B J and Fomalont E B 1984 {\it Astronomical J.} {\bf 89} 1650--57
\bibitem{Schweizer88}
Schweizer F and Seitzer 1988 {\it Astrophysical J.} {\bf 328} 88 
\bibitem{Schweizer80}
Schweizer F 1980 {\it Astrophysical J.} {\bf 237} 303--18
\bibitem{Matthews64}
Matthews T A, Morgan W W and Schmidt M 1964 {\it Astrophysical J.} {\bf 140} 35 
\bibitem{Goudfrooij2004}
Goudfrooij P, Gilmore D, Whitmore B C and Schweizer F 2004 {\it Astrophysical J.} {\bf 613} 121 
\bibitem{Mackie98}
Mackie G and Fabbiano G 1998 {\it Astrophysical J.} {\bf 115} 514
\bibitem{horellou2001}
Horellou C, Black J H, van Gorkom J H, Combes F, van der Hulst J M and Charmandaris V 2001
{\it A \& A} {\bf 376} 837--52
\bibitem{Bosma85}
Bosma A, Smith R M and Wellington K J 1985 {\it Mon. Not. R. Astron. Soc.} {\bf 212} 301
\bibitem{Grillmair99} 
Grillmair C J, Forbes D A, Brodie J P and Elson R A W 1999 {\it Astronomical J.} {\bf 117} 167--80
\bibitem{Lanz2010}  
Lanz L, Jones C, Forman W R, Ashby M L N, Kraft R and Hickox R 2010 {\it Astrophysical J.} {\bf 721} 1702--13
\bibitem{Temi2005}
Temi P, Mathews W G and Brighenti F 2005 {\it Astrophysical J.} {\bf 622} 235--43
\bibitem{Dale2007}
Dale D A {\it et al.} 2007 {\it A \& A} {\bf 655} 863--84
\bibitem{Draine2007}
Draine B T {\it et al.} 2007 {\it Astrophysical J.} {\bf 663} 866--94
\bibitem{kim2003}
Kim D W and Fabbiano G 2003 {\it Astrophysical J.} {\bf 586} 826
\bibitem{feigelson1995}
Feigelson E D, Laurent-Muehleisen S A,  Kollgaard R I and Fomalont E B 1995 {\it Astrophy. J. Lett.} {\bf 449} 149	
\bibitem{fabbiano1992}
Fabbiano G, Kim D W and Trinchieri G 1992 {\it Astrophysical J. Suppl. Ser.} {\bf 80}  531--644
\bibitem{hopkins2010}
Hopkins P F and Quataert E 2010 {\it Mon. Not. R. Astron. Soc.} {\bf 407} 1529
\bibitem{siringo2009}
Siringo G {\it et al.} 2009 {\it A \& A} D {\bf 497} 945--62
\bibitem{gusten2006}
G\"usten R, Nyman L A, Schilke P, Menten K, Cesarsky C and Booth R 2006 {\it A \& A} {\bf 497} 13
\bibitem{kovacs2008}
Kov\'{a}cs A 2008 {\it SPIE Proc.} {\bf 7020} 45K
\bibitem{wright2010}
Wright E L {\it et al.} 2010 {\it Astronomical J.} {\bf 140} 1868--81
\bibitem{jarrett2012}
Jarrett T H{\it et al.} 2012 {\it Astronomical J.} {\bf 144} 68
\bibitem{jarrett2013}
Jarrett T H{\it et al.} 2013 {\it Astronomical J.} {\bf 145} 1--34
\bibitem{cutri2012}
Cutri R M {\it et al.} 2012 {\it WISE All-Sky Data Release Products Explanatory Supplement} {\bf 2012wise.rept.}
\bibitem{peng2002}
Peng C Y, Ho L C, Impey C D, and Rix H 2002 {\it Astronomical J.} {\bf 124} 226
\bibitem{galametz2012}
Galametz M {\it et al.} 2012 {\it Mon. Not. R. Astron. Soc.} {\bf 425} 763 
\bibitem{galametz2014}
Galametz M {\it et al.} 2014 {\it Mon. Not. R. Astron. Soc.} {\bf 439} 2542
\end{thebibliography}
\end{document}